\documentclass[twocolumn,prl,a4paper,superscriptaddress,showpacs,showkey,nofootinbib]{revtex4}
\usepackage{amsfonts,amsmath,amssymb,graphicx,epsfig}

\def\chemin{./}

\newcommand{\pic}[3]{\includegraphics[clip=true,width=#2 \linewidth,angle=#3]{\chemin #1}}

\newcommand{\figcap}[2]{\caption{\label{#1}\em #2}}

\newcommand{\bmini}[1]{\begin{minipage}{#1 \linewidth}}
\newcommand{\emini}{\end{minipage}}

\begin{document}
\title{Influence of pore-scale disorder on viscous fingering during drainage}

\author{Renaud Toussaint}
\affiliation{Department of Physics, University of Oslo, POBox 1048 Blindern, N-0316 Oslo, Norway}

\author{Grunde L\o voll}
\affiliation{Department of Physics, University of Oslo, POBox 1048 Blindern, N-0316 Oslo, Norway}

\author{Yves M{\'e}heust}
\affiliation{Department of Physics, NTNU Trondheim, N-7491 Trondheim, Norway}

\author{Knut J\o rgen M\aa l\o y} \affiliation{Department of Physics, University of
Oslo, POBox 1048 Blindern, N-0316 Oslo, Norway} 
\author{Jean Schmittbuhl} \affiliation{Laboratoire de
G\'eologie, \'Ecole Normale Sup\'erieure, 24 rue Lhomond, F-75231 Paris, France}
\date{\today}

\begin{abstract}
We study viscous fingering during drainage experiments in linear  Hele-Shaw
cells filled with a random porous medium. The central zone of the cell is found to be statistically more occupied than the average, 
and to have a lateral width of 40\% of the system width, 
irrespectively of the capillary number $Ca$.
A crossover length $w_f \propto Ca^{-1}$ separates lower scales where
the invader's fractal dimension $D\simeq1.83$ is identical to capillary fingering, and
larger scales where the dimension is found to be $D\simeq1.53$. The
lateral width and the large scale dimension are lower than the results for Diffusion Limited Aggregation,
 but can be explained in terms of Dielectric Breakdown Model. Indeed, we show
that when averaging over the quenched disorder in capillary thresholds, an
effective law $v\propto (\nabla P)^2$ relates the average interface growth
rate and the local pressure gradient.
\end{abstract}



\maketitle

Viscous fingering instabilities in immiscible two-fluid flows in
porous materials have been intensely studied over the past 50
years \cite{BenSimon86}, both because of their important role in oil recovery
processes, and as a paradigm of simple pattern forming system. Their
dynamics is controlled by the interplay between viscous, capillary and
gravity forces. 
The ratio of viscous forces to the capillary
ones at pore scale is quantified by the capillary number $Ca=\mu v_f
a^2 / (\gamma \kappa)$, where $a$ is the characteristic pore size, $v_f$ is the filtration velocity,
$\gamma$ the interfacial tension,
$\kappa$ the permeability of the cell, and $\mu$ the viscosity of the displaced fluid, supposed here much larger than the viscosity of the invading one.

\begin{figure}
  \pic{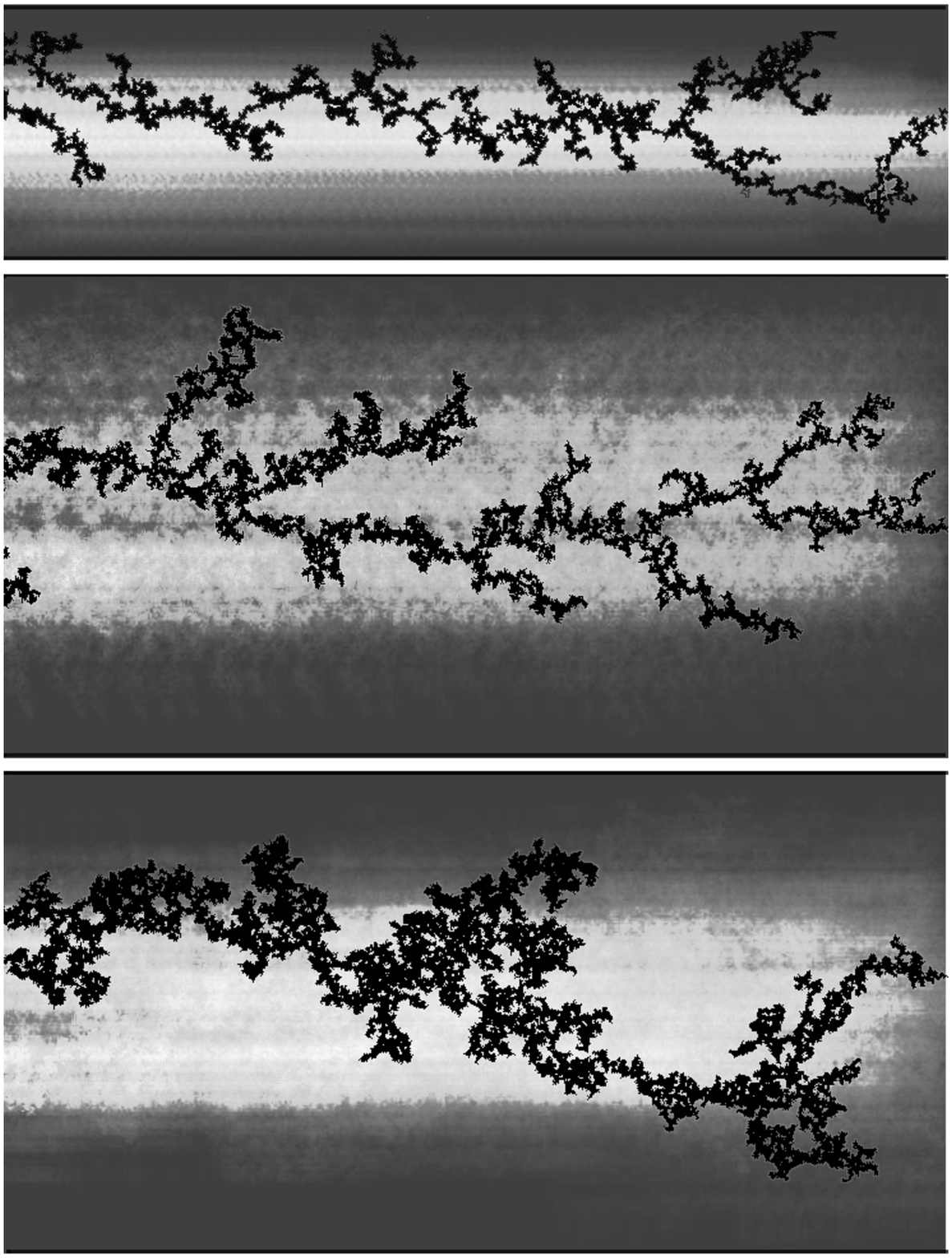}{0.6}{270} \figcap{fig_observations}{ Invasion clusters
    on thresholded images at capillary numbers ${\rm Ca} = 0.06$ {\bf
      (a)} and $0.22$ {\bf (b,c)}, for $W/a=210$ {\bf
      (a,b)}and $110$ {\bf(c)}, with
  displayed  system lateral boundaries. The superimposed gray map shows the occupancy probability $\pi(x,z)$ of the invader, in a moving reference frame attached to the most advanced invasion tip and to the lateral boundaries.}
\end{figure}

There is a strong analogy between viscous fingering in porous media and Diffusion
Limited Aggregation (DLA), as was first pointed out by Paterson
\cite{Paterson84}. Indeed, both processes of DLA and viscous fingering in empty Hele-Shaw cells belong to the family of Laplacian
Growth Models, i.e. obey the Laplacian growth equation $\nabla ^2 P =
0 $, with an interfacial growth rate $v \propto -\nabla P$, where P is
the diffusing field, i.e. the probability density of random walkers in
DLA, or the pressure in viscous fingering. Despite differences as
respectively a stochastic and deterministic growth, and boundary
conditions as respectively $P=0$ or $P=-\gamma/r$ with $r$ the
interfacial curvature, it is admitted that these processes belong
to the same universality class \cite{Paterson84,Stepanov01,Levermann04}. 
In radial geometry these
processes lead to fractal structures of dimensions $D=1.70\pm0.03$
\cite{Sharon03}, $1.713\pm0.0003$ \cite{Davidovitch00}, and $1.7$
\cite{Levermann04} respectively in viscous fingering in empty
Hele-Shaw cells, DLA, and numerical solutions of deterministic
Laplacian growth. The two numerical models have been reexplored recently using stochastic conformal
mapping theory \cite{Stepanov01,Davidovitch00,Levermann04}.  However, in Hele-Shaw cells filled with
disordered porous materials similar to the one used here, a lower
fractal dimension $D=1.58\pm0.09$ has been measured
\cite{Hinrichsen89}.

In straight channels, DLA gives rise to fractal
structures of dimension 1.71, occupying on average a lateral fraction
$\lambda=0.62$ of the system width $W$ \cite{Arneodo96}. Viscous
fingering in empty Hele-Shaw cell converge towards the Saffman Taylor
(ST) solution \cite{Saffman58}, with a uniformly propagating fingerlike
interface covering a fraction $\lambda=0.5$ of the system width at large capillary numbers
\cite{Saffman58,Tabeling87}, selected by the interfacial tension
\cite{McLean81}. 

 In the system we study, the cell is
filled with a disordered porous medium,
 and the non-wetting invader of low viscosity shows a
branched structure that depends on $Ca$ (Fig.~\ref{fig_observations}). We will show theoretically
that if indeed at high capillary number, the process is well described by DLA as often suggested \cite{Paterson84}, there is also
at intermediate $Ca$ a regime where the flow in {\it random} porous media  is better described by another Laplacian model, namely a
Dielectric Breakdown Model (DBM) with $\eta=2$ -- the interfacial
growth rate is $v \propto (\nabla P)^\eta$ in DBM, $\eta=1$ corresponding to DLA --.
 Our conclusion is supported by both experimental evidence based on a characterization of the 
large- and small-scale geometry of the invader,
and theoretical arguments based on averaging the capillary forces contribution to the interface growth rate over 
the pore scale randomness.
The vanishing capillary number limit of our system corresponds to
capillary fingering, where pores are invaded one by one, forming a
propagating front filling the whole system, leaving behind isolated
clusters of viscous fluid. The invader's fractal dimension is $D=1.83$
\cite{Lenormand85}, theoretically explained by the invasion
percolation model \cite{Chandler82}.  At moderate but finite capillary
numbers under interest here, we will show that small scales still
correspond to invasion percolation, up to a crossover scale
$w_f/a\propto Ca^{-1}$ characterizing the maximum size of
isolated clusters of defending fluid, above which the system
geometry can be described by the DBM with $\eta=2$.
\begin{figure}
  \pic{nfig2.eps}{0.8}{0} \figcap{phiz-nz}{Scaled invader's density $n(z)/n_{Ca}$ as function
of $2z/W$, distance to the most
    advanced tip scaled by the system's halfwidth, for three system
    sizes and four capillary numbers. The lines are the experimentally determined cumulative growth probability, and the theoretical Saffman Taylor solution for a single finger occupying a lateral fraction $lambda=0.4$ of the system.}
\end{figure}

We study viscous fingering processes in linear Hele-Shaw cells of
thickness $a=1$mm filled at 38\% with a monolayer of randomly located
immobile glass beads of diameter $a$, between which air displaces a
solution of 90\% glycerol - 10\% water of much larger viscosity
$\mu=0.165$ Pa$\cdot$s, wetting the beads and walls of the cell, i.e.
in drainage conditions.  
The interfacial tension and the permeability of the cell are respectively $\gamma=0.064$ N$\cdot$m$^{-1}$ and $\kappa=0.00166\pm0.00017$ mm$^2$.
We investigate regimes ranging from capillary to viscous fingering
($0.01<Ca<0.5$), in cells with impermeable lateral walls and dimensions $W \times L \times a$, with widths
perpendicularly to the flow direction $W/a=110$, $215$ and $430$, and a length $L/a=840$. 
The cell is set horizontally, so that gravity is irrelevant.
A constant filtration rate of water-glycerol is
ensured by a controlled gravity-driven pump. 

 Pictures of the flow pattern are taken from the top, and treated to extract the
invading air cluster (with pixels of size 0.55$a$), as the black clusters in 
Fig.~\ref{fig_observations}. In ref.~\cite{Lovoll04}, we have shown that 
the invasion process is stationary, up to fluctuations arising from the disorder in pore geometries.
To extract the underlying average stationary behavior, all quantities
are then analyzed in the reference frame $(x,z)$ attached to the
lateral boundaries at $x=0$ and $x=W$, and to the foremost propagating tip at
$z=0$, z pointing against the flow direction
(this tip indeed propagates at a roughly constant speed $v_{tip}$ \cite{Lovoll04}).
Average quantities at
any position $(x,z)$ of the tip related frame, are defined using all stages and points of the invasion
process, excluding regions closer than $W/2$ from the inlet or outlet,
to avoid finite size effects.

The average occupancy map $\pi(x,z)$ is defined as \cite{Lovoll04}: for each time (or each picture), 
we assign the value $1$ to the coordinate $(x,z)$ if air is present there, $0$ otherwise.
$\pi(x,z)$ obtained as the time average of such occupancy function, is displayed as graymap in Fig.~\ref{fig_observations}.

Next we compute the 
average number of occupied pores per unit length at a distance $z$ behind the tip, $n(z)$, which is related to 
$\pi$ as  
$n(z) = (1/a^2)\: \int_0^W \pi(x,z)\, dx$.
We show in Fig.~\ref{phiz-nz} a data collapse for different capillary numbers $Ca$ and system widths $W$, $n(z)/n_{Ca}=\Phi(2z/W)$, where $n_{Ca}=(W/a^2) v_f /v_{tip}$ \cite{Lovoll04}. $\Phi$ is a function increasing from $0$ at $z=0$ towards $1$ at $z=+\infty$, as granted by conservation of the displaced fluid for a statistically stationary process  \cite{Lovoll04}. $\Phi$ evaluated in  Fig.~\ref{phiz-nz} is a cumulative growth probability defined in Ref.~\cite{Lovoll04}, and is obtained as an average over all experiments and sizes.

We also characterize the lateral structure of the invader in the frozen zone, $z>W$, where less than 10\% of the invasion activity takes place 
since $\Phi(2)>0.9$. We 
 define over this zone a distribution $\rho(x)= [W/(a^2\, n_{Ca})]\: \pi(x,\infty) = (v_\text{tip}/v_\text{f}) \: \pi(x,\infty)$, so that 
$\int_0^W \rho(x) dx/W \simeq 1$.
This quantity, presented in Fig.~\ref{half-occ-map}(a) for an average over five experiments at capillary
numbers $Ca=0.06$ and $0.22$ for $W/a=215$, and over four experiments with $0.06<Ca<0.22$ for $W/a=110$,
 is reasonably
independent of the capillary number and the system size, though the noise is larger at
highest speed. The fraction $\lambda$ of the system,
occupied by the invader at saturation is evaluated as in
\cite{Arneodo96}: $\lambda=1/\rho_{max}$, or alternatively
$\lambda=(x^+ -x^-)/W$, where $\rho(x^+)=\rho(x^-)=\rho_{max}/2$. Both
definitions lead to $\lambda \simeq 0.4\pm 0.02$ for the capillary
numbers probed, as shown in Fig.~\ref{half-occ-map}(a), which is
significantly smaller than the off-lattice DLA result $\lambda=0.62$
\cite{Arneodo96,Somfai03}.
\begin{figure} 
  \pic{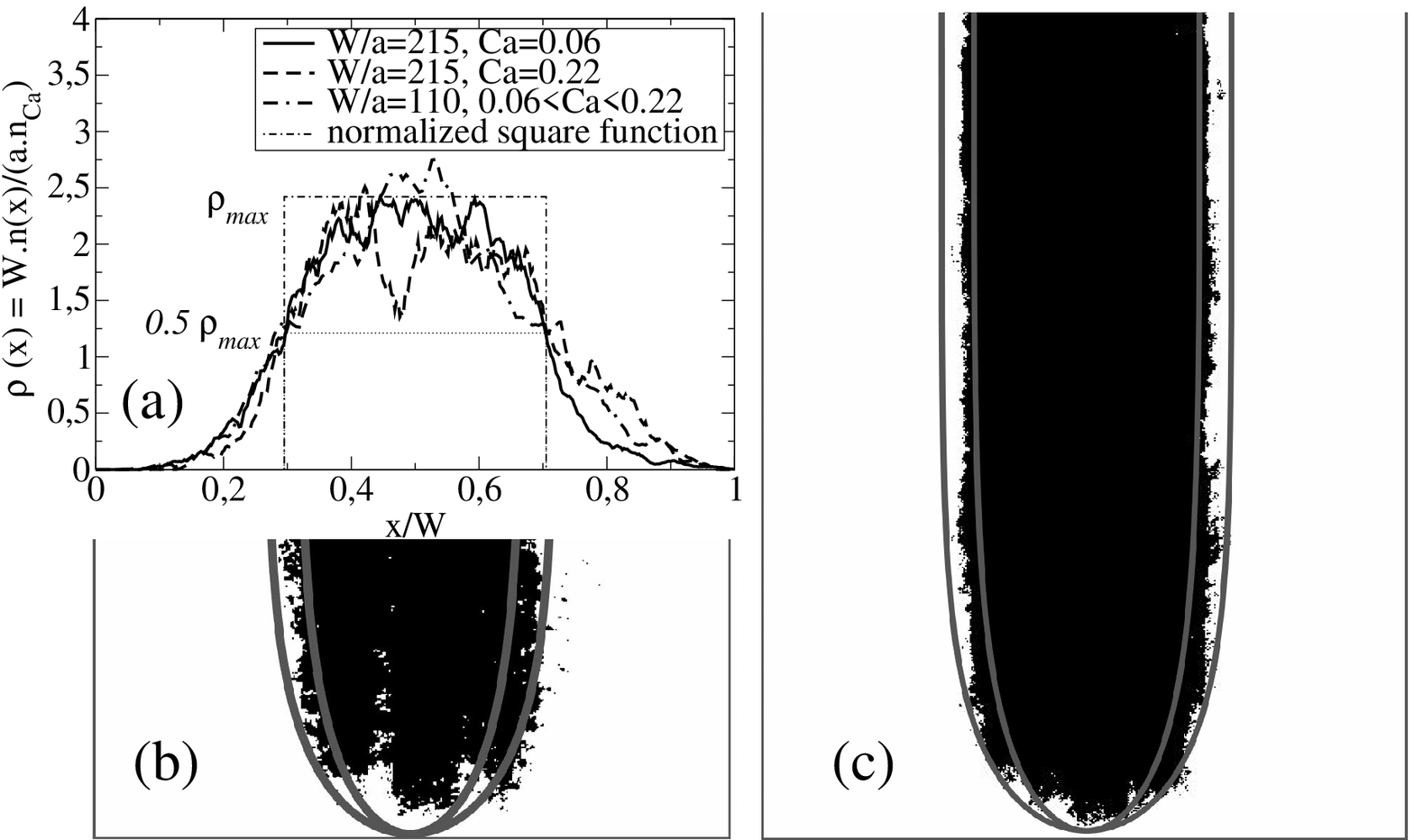}{0.9}{0} \figcap{half-occ-map}{(a) Normalized
    occupation density $\rho(x)$,
    with half-maximum reached over a width $0.4 W$.  (b,c) Average
    occupation density map of the invader thresholded at half-maximum, for a system size W/a=215,
    in the reference frame attached to the tip position, at $Ca=0.06$
    (b) and $0.22$ (c), compared to ST curves for $\lambda=0.35$
    and $0.45$.}
\end{figure}

The 2D occupancy map $\pi(x,z)$ itself, displayed as graymap in Fig.~\ref{fig_observations}, has a maximum
$\pi_\text{max}=\rho_\text{max} v_f/v_\text{tip}$, along a line at $(x=W/2,z>W)$.
Similarly to Arneodo's procedure \cite{Arneodo96}, we determine the support of $\pi>\pi_{max}/2$, displayed in 
Fig.~\ref{half-occ-map}(b) and (c), which corresponds to the most often
occupied region.
 Within the noise error, the shape of this region resembles
the theoretical ST curve corresponding to this $\lambda=0.4$ \cite{Saffman58} (gray lines in Fig.~\ref{half-occ-map}).
For such ST finger, this curve would also correspond to the scaled longitudinal density of invader $\Phi(z)$. 
$\Phi(z)$ determined for our experiments (continuous line in Fig.~\ref{phiz-nz}) and this theoretical ST shape (dashed line) also present some similarities.
Systematic deviations from the mathematical ST solution at $\lambda=0.4$ might nonetheless exist, since they have been seen between the DLA envelopes 
and the corresponding ST solution at $\lambda=0.62$ \cite{Somfai03}.

\begin{figure}
  \pic{nfig4.eps}{0.8}{0} \figcap{fractaldims}{ (a) Mass
    fractal dimension of the invasion cluster at various capillary
    numbers.  (b): Cross-over scale $w_f$ between capillary and
    viscous scales, as function of $Ca$.}
\end{figure}

The mass fractal dimension of the invasion clusters is analysed by
box-counting. $N(s)$ is the number of boxes of size $s$ to cover the invader.
Fig.~\ref{fractaldims} displays a normalized distribution $N(s)/N(a/Ca)$ as function
of $(s/a)\cdot Ca$ for various capillary numbers. By linear regression
of this collapsed log-log data, we find that $N(s)\sim
s^{-1.83\pm0.01}$ for small scales $s<a/Ca$, and $N(s)\sim
s^{-1.53\pm0.02}$ for larger scales.  The result can be explained
by the following approximations. The distribution of pore throats
sizes results in a distribution of capillary pressure thresholds $P_t$,
 $g(P_t)$, of characteristic width $W_t$.
Consider a box of size $w_f$ along the cluster boundary in the active zone,
 such that $w_f\cdot \nabla P_b = W_t$
where $\nabla P_b$ is a characteristic pressure gradient. At scales $s<w_f$,
 viscous pressure variations are lower than capillary threshold fluctuations, and
the most likely invaded pores correspond to the lowest random
thresholds, which corresponds to capillary fingering, thus leading to
$D=1.83$. Conversely, at larger sizes ($s>w_f$), the invasion activity is
 determined essentially by the spatial variations of $\nabla P_b$.
Assuming that $\nabla P_b$ scales as the imposed $\nabla
P(-\infty)\propto Ca$, i.e. neglecting the geometry variations between
different speeds, $w_f$ scales as $W_t/\nabla P_b
\propto a/Ca$, as confirmed by the data collapse in
Fig.~\ref{fractaldims}(a). $w_f$ can also be
determined experimentally as a characteristic branch width, since
capillary fingering leaves isolated clusters of trapped fluid, whereas
the large scale structure is branched. After removing all trapped
clusters, we determine $w_f$ as the average length of intersects of
the structure from cuts along x. Indeed,
Fig.~\ref{fractaldims}(b) is consistent with a scaling law $w_f/a \propto Ca^{-1}$, below a
saturation at large $Ca$. 

Eventually, we sketch a possible explanation for the width selection
$\lambda=0.4$ and the large scale fractal dimension $D=1.53\pm0.02$,
which are smaller than their counterparts in DLA, respectively 0.62
and 1.71. Neglecting the small scale permeability variations leads to
a Laplacian pressure field in the defending fluid. The boundary
condition for the pressure field is then $\nabla P(z=-\infty) = - \mu
v_f/ \kappa$, and $\nabla P(x=0,W) \cdot \hat{x} =0$ where $\hat{x}$ is the unit vector along $x$.
 The dynamics of
the process is then entirely controlled by the boundary condition
along the invading fluid, i.e. by the capillary pressure drop across the meniscus in the pore neck and the pressure gradient in the invaded fluid. For a given pressure difference at pore scale between the invading air at $P_0$, and the pressure $P_1$ in the glycerol-filled pore, we decompose $P_0-P_1=\Delta P_v+P_c$, where $\Delta P_v$ is a viscous pressure drop in the pore neck, and the capillary pressure drop is $P_c=\gamma/r+2\gamma/a$, where the in- and out-of-plane curvature of the interface are respectively $r$ and  $a/2$.
As a meniscus progresses between neighboring beads, its curvature goes
through a minimum $r_m$ in the pore neck. The meniscus will be able to pass the neck if the
 pressure drop $P_0-P_1$ exceeds the
threshold $P_t=P_c(r_m)$.  For the sake of simplicity, the probability
distribution of the thresholds $g(P_t)$ is considered flat, between $P_{min}$ and $P_{max}$, with
$W_t=P_{max}-P_{min}$ and
$g(P_t)=\theta(P_t-P_{min})\theta(P_{max}-P_t)/W_t$, where $\theta$ is the Heaviside function.
In the pure
capillary fingering limit $Ca\rightarrow 0$, the pressure field $P$ is
homogeneous in the defending fluid, and a pore is invaded when $P_0-P$
reaches the minimum threshold along the boundary, close to $P_{min}$.
At higher capillary number, 
we
want to relate the invasion rate to the
local capillary threshold, and to the pressure $P_1$ in the
liquid-filled pore nearest to the interface. If $P_0-P_1<P_t$, the meniscus adjusts reversibly in
the pore neck, and the next pore is not invaded. Conversely, if
$P_0-P_1>P_t$, the pore will be invaded,
and most of the invasion time is spent in the thinnest region of the pore neck. A
characteristic interface velocity can be evaluated by the Washburn
equation \cite{Washburn} at this point: $v\sim -(2\kappa / \mu a) (P_0-P_1-P_t)
\theta(P_0-P_1-P_t)$, where the heaviside function results from a zero
invasion velocity if the pore is not invaded. Hypothesizing that
only the average growth rate controls the process, independently of
the particular realization of random thresholds, the growth rate averaged over all
possible pore neck configurations, is
\begin{eqnarray}
\lefteqn{<v>= \int  \frac{-2\kappa}{\mu a} (P_0-P_1-P_t) \theta(P_0-P_1-P_t) g(P_t) dP_t}&&\nonumber\\
&=& -(\kappa / \mu a)\theta(P_0-P_1-P_{min}) \times \label{eqinvasion}\\
&& \{[(P_0-P_1-P_{min})^2/W_t] \theta[P_{max}-(P_0-P_1)] + \nonumber\\
&& 2[P_0-P_1-(P_{min}+P_{max})/2] \theta[P_0-P_1-P_{max}]\}\nonumber.
\end{eqnarray}
At moderate capillary numbers, such as $P_0-P_1<P_{max}$, if we assume that the capillary pressure drop is around $P_c=P_{min}$ when the invasion meniscus is at the entrance of the pore neck, we note that
$(P_0-P_{min}-P_1)/a=\Delta P_v/a \sim \nabla P / 2$, and Eq.~(\ref{eqinvasion}) implies that the
growth rate goes as $<v>=-a \kappa / (4 \mu W_t) (\nabla P)^2$.
This effective quadratic relationship between the average growth rate
and the local pressure gradient arises from the distribution
of capillary thresholds, and means that such invasion process should
be in the universality class of DBM with $\eta=2$, rather than DLA
(DBM, $\eta=1$). Indeed, in DBM
simulations in linear channels, $\lambda$ is a decreasing function of
$\eta$ (as in related deterministic problems, as viscous fingering in shear-thinning fluids \cite{Lindner02}, or $\eta$-model \cite{BenAmar95}), and Somfai et al. \cite{Somfai03} report $\lambda\simeq 0.62$
and $0.5$ for respectively $\eta=1$ and $1.5$, so that the observed
$\lambda=0.4$ is consistent with $\eta=2$.  The fractal dimension of
DBM is also a decreasing function of $\eta$, and $\eta=2$ corresponds $D=1.4\pm0.1$
\cite{Mathiesen02}, which is close to the observed $D =
1.53\pm0.02$ in our experiments.

Note that at high capillary numbers such that
locally $P_0-P_1 \gg P_{max}$, the threshold fluctuations are not felt by
the interface, and Eq.~(\ref{eqinvasion}) leads to $<v_{inv}> = - (\kappa / \mu)
 \nabla P$, which would correspond to a classic DLA process. We have checked 
by numerically solving the Laplace equation with the experimental clusters as boundaries that all experiments performed here were at moderate enough capillary number to have $P_0-P_1<P_{max}$ all along the boundary \cite{Lovoll04}, i.e. the quadratic law $<v>=-a \kappa / (4 \mu W_t) (\nabla P)^2$ is expected to hold. 

Even at moderate $Ca$, deviations from the DBM model with $\eta=2$ could be observed for significantly non-flat
distribution of the capillary thresholds in the random porous medium, for which Eq.(\ref{eqinvasion}) would lead to a more
complicated dependence of the growth rate $v$ on $\nabla P$, reflecting the details
of this distribution, and not simply a power-law effective
relationship.  It would be interesting in future work to explore numerically and experimentally the
detailed effect of non-flat capillary threshold distributions on the selected
fractal dimension, average width occupied in the system, and total
displaced mass $n_{Ca} (Ca)$ (reported in \cite{Lovoll04} for the present work), to extract the influence of the disorder
on the best capillary number to select in order to maximize the
efficiency of the extraction process.

We acknowledge with pleasure fruitful discussions with E.~G. Flekk{\o}y, A. Lindner, A. Hansen and E. Somfai. This  work  was  supported  by  the  CNRS/NFR PICS program,  and  the NFR Petromax program.

\bibliographystyle{apsrev}


\end{document}